\documentclass[aps]{revtex4}

\begin{document}

\title{Analytical soft-core potentials for macromolecular fluids and mixtures}

\author{G.\ Yatsenko, E.\ J.\ Sambriski, M.\ A.\ Nemirovskaya, M.\ Guenza}

\affiliation{Institute of Theoretical Science, Department of Chemistry,
University of Oregon, Eugene, OR 97403}

\begin{abstract}
An analytical description of polymer melts and their mixtures as liquids of
interacting soft colloidal particles is obtained from liquid-state theory. The
derived center-of-mass pair correlation functions with no adjustable parameters
reproduce those computed from united atom molecular dynamics simulations.
The coarse-grained description correctly bridges micro- and mesoscopic fluid
properties. Molecular dynamics simulations of soft colloidal particles
interacting through the calculated effective pair potentials are consistent with
data from microscale simulations and analytical formulas.
\end{abstract}

\maketitle

The formulation of an accurate mesoscopic description of macromolecular fluids
has been a longstanding goal in polymer physics. Experimentally-relevant polymer
dynamics span a wide range of timescales, for which large-scale, long-time
properties still depend strongly on the local molecular structure
\cite{Frenkel}. Pertinent information on structure and dynamics of
polymer liquids has been gained from united atom (UA) molecular dynamics (MD)
simulations. However, the MD computational time increases as the squared number
of interacting units, and the latter has to be large to approximate the
thermodynamic limit, rendering an all-atom simulation of long-time polymer
dynamics a prohibitive task. One strategy devised to overcome this problem is to
renormalize the liquid structure and dynamics using effective-unit
coarse-grained descriptions \cite{Frenkel}. Specifically, polymers can be
described mesoscopically as soft interpenetrating spheres having the overall
size of the polymer, i.e., the radius of gyration $R_g$. However, to correctly
perform the renormalization procedure, a theoretical framework that bridges
properly different lengthscales of interest is needed. Phenomenological
mesoscopic potentials were implemented by Dautenhahn and Hall,
and later on by Murat and Kremers, to describe polymer melts and blends
\cite{Hall,Kremer}. Hansen and coworkers have recently developed a
rigorous numerical description of polymer solutions as liquids of soft
interacting colloidal particles \cite{Hansen}.

In this Letter we start from first-principles liquid-state theory and derive an
analytical form of center-of-mass (c.o.m.) pair correlation functions, from
which the effective pair soft-core potential acting between molecules in polymer
liquids (melts) and their mixtures (blends) is obtained. The c.o.m.\ pair
correlation functions reproduce mesoscale liquid structures obtained from UA-MD
simulations \cite{Grest,Grest1,Grest2} without adjustable parameters. Test
systems are polymer melts with different architecture, local semiflexibility,
and degree of polymerization (Table I), as well as their mixtures (Table II).
Finally, the mesoscopic potential derived by an inversion procedure is used in
MD simulations of soft colloidal particles, which reproduce the liquid structure
at the level of c.o.m.\ pair correlation functions.

The renormalized pair interaction potential is a function of the c.o.m. total
pair correlation function $h(r)$. In reciprocal space,
$h(k)=[{\omega^{mC}(k)}/{\omega^{mm}(k)}]^2 h^{mm}(k)$, after a procedure
devised by Krakoviack, Hansen, and Louis \cite{Krakov2002}. Here,
$\omega^{mC}(k)$ is the intrachain monomer distribution about the
c.o.m. Also, $h^{mm}(k)$ and $\omega^{mm}(k)$ are the monomer-monomer
intermolecular total pair and intramolecular correlation functions,
respectively. We assume a Gaussian description for intramolecular form factors:
$\omega^{mm}(k)=N/(1+k^2R_g^2/2)$, where $N$ is the number of ``monomeric''
units of length $\sigma$ and $R_g=\sqrt{N/6}\sigma$ is the overall chain size,
and $\omega^{mC}(k)=Ne^{-k^2R_g^2/6}$ \cite{yamakawa}. For $h^{mm}(k)$, we use
the thread-limit PRISM description \cite{PRISM}, in which the chain is treated
as an infinite thread of vanishing thickness.  Since this model correctly
reproduces the appearance of correlation hole effects on the scale of $R_g$, a
distinguishing feature of macromolecular fluids, liquid properties at the
mesoscale of interest should be described well. Now,
$h^{mm}(k)=4\pi\xi_{\rho}' \left[\xi_{\rho}^2/(1+k^2\xi_{\rho}^2)-
\xi_c^2/(1+k^2\xi_c^2) \right]$, where $\xi_c=R_g/\sqrt{2}$ is the correlation
hole lengthscale. The density fluctuation length is defined as
$\xi_{\rho}^{-1}=\xi_c^{-1}+\xi_{\rho}'^{-1}$ with
$\xi_{\rho}'=R_g/(2\pi\rho_{ch}^*)$, and
$\rho_{ch}^*=\rho_{ch}R_g^3$, where $\rho_{ch}$ is the chain number density. In
real space, we find
\begin{eqnarray}
h(r) & = &
\frac{3}{4}\sqrt{\frac{3}{\pi}}\frac{\xi_{\rho}'}{R_g}
\left(1-\frac{\xi_c^2}{\xi_{\rho}^2}\right)
e^{-3r^2/(4R_g^2)}
-\frac{1}{2}\frac{\xi_{\rho}'}{r}
\left(1-\frac{\xi_c^2}{\xi_{\rho}^2}\right)^2
e^{R_g^2/(3\xi_{\rho}^2)}\nonumber \\
&& \times\left[
e^{r/\xi_{\rho}}\mbox{erfc}
\left(\frac{R_g}{\xi_{\rho}\sqrt{3}}+\frac{r\sqrt{3}}{2R_g}\right)
-e^{-r/\xi_{\rho}}\mbox{erfc}
\left(\frac{R_g}{\xi_{\rho}\sqrt{3}}-\frac{r\sqrt{3}}{2R_g}\right)
\right] \ ,
\label{gccR}
\end{eqnarray}
where $\mbox{erfc}(x)$ is the complementary error function. This expression
satisfies the condition $h(r) \ge -1$, spanning melt to dilute solution
densities up to $\rho_{ch}^* \ge 0.03$. Eq.\ (\ref{gccR}) correctly recovers the
isothermal compressibility $\kappa_T$ \cite{HansenMcd} related to the
$k\rightarrow 0$ limit of the mesoscale static structure factor as
$S(0)=1+\rho_{ch} h(0)=(\xi_{\rho}/\xi_c)^2=\rho_{ch} k_B T \kappa_T$.

Our total correlation function $h(r)$ effectively maps the polymeric liquid onto
a fluid of soft interpenetrating colloids of radius $R_g$ and density
$\rho_{ch}$. Translated into the colloidal-particle framework, Eq.\ (\ref{gccR})
is a function of the reduced variables $\tilde{r}\equiv r/R_g$,
$\tilde{\xi}_{\rho}'=\xi_{\rho}'/R_g \equiv (2 \pi \rho^*_{ch})^{-1}$, and
$ \tilde{\xi}_{\rho}=\xi_{\rho}/R_g \equiv
\left[\sqrt{2}(1+\pi\sqrt{2}\rho_{ch}^*)\right]^{-1}$.
Then, Eq.\ (\ref{gccR}) reduces to $h(r)\equiv h(\tilde{r},\tilde{\xi}_{\rho})$,
where we define
\begin{eqnarray}
h(\tilde{r},\tilde{\xi}_{\lambda}) & = &
\frac{3}{4}\sqrt{\frac{3}{\pi}}\tilde{\xi}_{\rho}'
\left(1-\frac{1}{2\tilde{\xi}_{\lambda}^2}\right)
e^{-3\tilde{r}^2/4}
- \frac{1}{2}\frac{\tilde{\xi}_{\rho}'}{\tilde{r}}
\left(1-\frac{1}{2\tilde{\xi}_{\lambda}^2}\right)^2
e^{1/(3\tilde{\xi}_{\lambda}^2)}\nonumber \\
&&\times\left[
e^{\tilde{r}/\tilde{\xi}_{\lambda}}\mbox{erfc}
\left(\frac{1}{\tilde{\xi}_{\lambda}\sqrt{3}}+\frac{\tilde{r}\sqrt{3}}{2}\right)
- e^{-\tilde{r}/\tilde{\xi}_{\lambda}}\mbox{erfc}
\left(\frac{1}{\tilde{\xi}_{\lambda}\sqrt{3}}-\frac{\tilde{r}\sqrt{3}}{2}\right)
\right] \ .
\label{univIcc}
\end{eqnarray}

We tested Eq.\ (\ref{gccR}) against UA-MD simulations for polymer melts of
polyethylene (PE) for increasing degree of polymerization ($N=44$, $66$, $96$)
\cite{Grest} as well as different local structure and flexibility, including
syndiotactic (sPP), isotactic (iPP) and head-to-head (hhPP) polypropylenes \cite
{Grest1,Grest2}. Input parameters to the theory were site number density $\rho$,
temperature $T$, and $R_g$ from Table I. Our analytical expression agrees well
with $h(r)$ from simulations within statistical error (Figs.\ 1 and 2).
Specifically, for PE melts it reproduces the tendency for chains to
interpenetrate more efficiently with increasing length, an effect due to the
fractal dimension of polymer chains. For slightly branched polymer melts, the
number of interpolymer contacts becomes larger with increasing polymer stiffness
(increasing $R_g$ at constant $N$). Good agreement between theory and computer
simulations is found also in reciprocal space, where the analytical Fourier
transform of Eq.\ (\ref{gccR}) reproduces simulation data in the entire range
for $k \ge R_g^{-1}$ (Fig.\ 3).

A simplified form of Eq.\ (\ref{gccR}) can be derived in the limit of long
polymer chains ($\tilde{\xi}_{\rho} \rightarrow 0$) as is implicitly assumed by
the use of Gaussian form factors, for which we find
\begin{eqnarray}
h(\tilde{r},\tilde{\xi}_\rho) & \approx &
- \frac{39}{16}\sqrt{\frac{3}{\pi}}
\tilde{\xi}_{\rho}\left(1+\sqrt{2}\tilde{\xi}_{\rho}\right)
\left(1-\frac{9\tilde{r}^2}{26}\right)
e^{-3\tilde{r}^2/4}
\ ,
\label{apgccR}
\end{eqnarray}
where $h(\tilde{r},\tilde{\xi}_{\rho})/\tilde{\xi}_{\rho}$ is a universal
function of the reduced distance $\tilde{r}$ up to the second-order correction
in $\tilde{\xi}_{\rho}$. Eq.\ (\ref{apgccR}) is a good approximation for Eq.\
(\ref{gccR}): results obtained with the two expressions cannot be distinguished
in Fig.\ 1. Both equations recover the correct trend of greater chain
interpenetration with increasing $R_g$ and/or liquid density as
$h(0)\propto (\rho_{ch}^*)^{-1}$.

From Eq.\ (\ref{gccR}), the effective potential is calculated self-consistently
using an inversion procedure based on the HNC approximation,
$\beta v(r)=h(r)-\mbox{ln}[h(r)+1]-c(r)$, where $\beta=1/k_BT$. The direct
correlation function $c(r)$ \cite{HansenMcd} is obtained from the Fourier
transform of the Ornstein-Zernike equation $c(k)=h(k)/\left[1+\rho_{ch}h(k)
\right]$. While the HNC closure is known to work well for dilute colloidal
systems, its applicability is questionable in dense systems where many-body
interactions should be important \cite{HansenMcd}. However, the pair HNC
interaction potential works well for the mesoscopically-renormalized polymer
melts investigated here. We performed constant temperature MD simulations of a
liquid of soft colloidal particles of radius $R_g$ interacting through the
derived pair potential, for chains of $N=44$, $66$, and $96$ units.
Corresponding effective potentials $\beta v(r)$ as functions of the normalized
distance $r/R_g$ differ slightly for the three cases. The potential for $N=44$
is shown in the inset of Fig.\ 1. Mesoscale simulations recover the liquid
structure obtained in UA-MD simulations in both real (Fig.\ 1) and
reciprocal space (Fig.\ 3).

We extended the formalism just presented to treat binary polymer mixtures. The
homopolymer species are $A$ and $B$, characterized by radii of gyration $R_{gA}$
and $R_{gB}$, with number of chain units $N_A$ and $N_B$, and unit lengths
$\sigma_A=\sqrt{6/N_A}R_{gA}$ and $\sigma_B=\sqrt{6/N_B}R_{gB}$. The volume
fraction of component $A$ is given by $\phi$ and $\gamma=
\sigma_B/\sigma_A$ represents the mismatch in local chain semiflexibility. For a
generic pair $\{\alpha,\beta \in (A,B)\}$, we have
$h_{\alpha \beta}(k)=\left[{\omega^{mC}_{\alpha}(k)\omega^{mC}_{\beta}}(k)/
({\omega^{mm}_{\alpha}(k)\omega^{mm}_{\beta}}(k))\right]^2 h^{mm}_{\alpha
\beta}(k)$. Chains are assumed to obey Gaussian intramolecular distributions
with $\omega^{mC}(k)$ and $\omega^{mm}(k)$ defined as in the melt case. The
$h^{mm}_{\alpha \beta}(k)$ follow the PRISM-blend thread model described by
Tang and Schweizer \cite{TangSch}, and is extended here to include asymmetries
in local chemical structure and flexibility ($\sigma_A \leq \sigma_B$). The
advantage of this approach is that no closure approximations need to be invoked,
while the miscibility parameter $\chi$ enters as an input to the theory
\cite{Balsara}. This allows us to describe the renormalized structure and
potential for systems following either upper or lower critical solution
temperature phase diagrams within one theoretical framework.

On applying inverse Fourier transforms, the blend $h_{\alpha \beta}(k)$ in real
space are given by
\begin{eqnarray}
h_{AA}(r) & = & \frac{1-\phi}{\phi}I^{\phi}_{AA}(r)+
\gamma^{2}I^{\rho}_{AA}(r) \ , \ \
h_{BB}(r)=\frac{\phi}{1-\phi}I^{\phi}_{BB}(r)+
\frac{1}{\gamma^{2}}I^{\rho}_{BB}(r) \ , \nonumber \\
h_{AB}(r)& = & -I^{\phi}_{AB}(r)+
I^{\rho}_{AB}(r) \ ,
\label{hhh}
\end{eqnarray}
with
\begin{eqnarray}
I^{\lambda}_{\alpha\beta}(r)&=&
\frac{3}{4}\sqrt{\frac{3}{\pi}}\frac{\xi_{\rho}'}
{R_{g\alpha\beta}}\vartheta_{\alpha\beta 1}
\left(1-\frac{\xi_{c\alpha\beta}^2}{\xi_{\lambda}^2}\right)
e^{-3r^2/(4R_{g\alpha\beta}^2)}
- \frac{1}{2}\frac{\xi_{\rho}'}{r}\vartheta_{\alpha\beta 2}
\left(1-\frac{\xi_{c\alpha\beta}^2}{\xi_{\lambda}^2}\right)^2
e^{R_{g\alpha\beta}^2/(3\xi_{\lambda}^2)}\nonumber \\
&& \times\left[
e^{r/\xi_{\lambda}}\mbox{erfc}
\left(\frac{R_{g\alpha\beta}}{\xi_{\lambda} \sqrt{3}}
+ \frac{r\sqrt{3}}{2R_{g\alpha\beta}}\right)
- e^{-r/\xi_{\lambda}}\mbox{erfc}
\left(\frac{R_{g\alpha\beta}}{\xi_{\lambda} \sqrt{3}}
- \frac{r\sqrt{3}}{2R_{g\alpha\beta}}\right) \right] \ .
\label{Ta}
\end{eqnarray}
Here $\xi_\lambda$ for $\{\lambda \in (\rho,\phi)\}$ identifies the lengthscale
for density or concentration fluctuation correlations. The concentration
fluctuation length $\xi_{\phi} = \sigma_{AB}/\sqrt{24(\chi_s-\chi)\phi(1-\phi)}$
diverges at the spinodal temperature where $\chi=\chi_s$. The average segment
length is $\sigma_{AB}^2=\phi \sigma_B^2+(1-\phi) \sigma_A^2$,
while $R_{g \alpha \beta}\equiv \sqrt{(R_{g\alpha}^2
+ R_{g\beta}^2)/2}=\xi_{c\alpha \beta}\sqrt{2}$, with $\xi_{c\alpha \beta}$
being the average correlation hole length. Also, $\vartheta_{\alpha\beta1}=
\left[1-\xi_{c\alpha\alpha}^{2}\xi_{c\beta\beta}^{2}/
\Big({\xi_{c\alpha\beta}^2\xi_{\lambda}^2}\Big)\right]/
\left[1-\xi_{c\alpha\beta}^2/\xi_{\lambda}^2\right]$ and
$\vartheta_{\alpha\beta2}=
\left[\Big(1-\xi_{c\alpha\alpha}^{2}/\xi_{\lambda}^{2}\Big)
\Big(1-\xi_{c\beta\beta}^{2}/\xi_{\lambda}^2\Big)\right] /
\left[1-\xi_{c\alpha\beta}^{2}/\xi_{\lambda}^2\right]^2$. The self terms in the
density fluctuation contributions $I^{\rho}_{\alpha\alpha}(r)$ are formally
identical to the total pair correlation function in the melt. For a totally
symmetric blend ($N_A=N_B$ with $\sigma_A=\sigma_B$), Eqs.\ (\ref{hhh},\ref{Ta})
correctly recover the melt equation in the athermal limit where enthalpic
effects are negligible.

We tested Eqs.\ (\ref{hhh},\ref{Ta}) against UA-MD simulations of blends
\cite{Grest1,Grest2} with polymer components having different degree of
polymerization $N$, local semiflexibility/branching $\sigma$, and volume
fraction $\phi$. By convention, the $B$-component is assumed to be the stiffest.
In Fig.\ 4, we report only two representative examples (see Table II). For a
thermal $\phi$-symmetric mixture of linear (PE) and slightly branched (hhPP)
molecules, the theory agrees well with simulations ($\chi=0.0016$ Ref.\
\cite{Balsara} and $\chi_s=0.0211$). The stiffest component (PE) shows a higher
number of intermolecular contacts than the flexible one (hhPP). A weak
self-clustering of the most flexible species is observed at a distance
comparable to the polymer size, while the number of $AB$ contacts is still large
at this temperature, an indication that the system is far from its phase
transition. Similar effects are shown for the second system in Fig.\ 4, which is
a thermal $\phi$-asymmetric blend of slightly branched (iPP) and linear (PE)
molecules ($\chi=0.005$ Ref.\ \cite{Grest1} and $\chi_s=0.0281$). In asymmetric
mixtures, the minority species (iPP) presents an enhanced clustering, in
agreement with experimental data and simulations \cite{Grest1,Grest2}.

The blend pair correlation functions can be mapped onto a system of interacting
soft colloids of dimension $R_{gA}$ and $R_{gB}$, with chain volume fraction
given by $\phi=\phi_{ch}$ and mismatch in chain size given by
$\gamma=R_{gB}/R_{gA}$. The analytical equations are considerably simplified for
symmetric mixtures where $\gamma=1$, and we confine our presentation to this
case \cite{BlndPpr}. For each pair interaction, Eq.\ (\ref{Ta}) reduces to Eq.\
(\ref{univIcc}): $I^{\lambda}(\tilde{r},\tilde{\xi}_{\lambda})\equiv
h(\tilde{r},\tilde{\xi}_{\lambda})$ where $\tilde{\xi}_{\lambda}$ for
$\{\lambda \in (\rho,\phi)\}$ and $\tilde{\xi}_{\phi}=\xi_{\phi}/R_{g} \equiv
1/\sqrt{2(1-\chi/\chi_s)}$. By combining Eqs.\ (\ref{hhh},\ref{Ta}) into density
and concentration fluctuation contributions following Bhatia-Thornton's
formalism \cite{HansenMcd}, we recover known expressions for a mixture of
symmetric colloidal particles (e.g., liquid alloys). For example, in the $k
\rightarrow 0$ limit, the correlation in number density $S^{\rho \rho}(0)=
(\xi_\rho/\xi_c)^2$ correctly recovers $\kappa_T$, while the concentration
fluctuation correlation reduces to the known formula $S^{\phi \phi}(0)=
\phi(1-\phi)/[1-2\phi(1-\phi)N\chi]$ upon introducing Flory's definition of
$\chi_s$. Finally, the cross term $S^{\rho \phi}(0)$ vanishes in agreement with
the theory of symmetric particle mixtures \cite{HansenMcd}. Consistency between
Eqs.\ (\ref{hhh},\ref{Ta}) and known properties of colloidal mixtures supports
our mapping of polymer blends onto a mixture of interacting soft colloidal
particles.

Summarizing, we report here the derivation of an analytical renormalized
description of polymer melts and blends as fluids of mesoscopic soft colloidal
particles. The related soft-core Gaussian potential explicitly bridges meso- and
microscale properties. The derived c.o.m.\ pair correlation functions
reproduce, in both {\it r-} and {\it k-}space, fluid structures obtained from
UA-MD simulations. Using the melt potential calculated from the HNC
approximation, we perform MD simulations of liquids comprised of soft colloidal
particles, from which the obtained pair correlation functions correctly recover
the liquid structure from UA-MD simulations, further supporting our renormalized
analytical description of polymer liquids.

We are grateful to G.\ S.\ Grest and E. Jaramillo for sharing UA-MD simulation
trajectories. We acknowledge support from the National Science Foundation
under grant DMR-0207949. Also, EJS acknowledges support from an NSF Graduate
Research Fellowship.

\pagebreak

TABLE I. Polyolefin melts.\\
\begin{tabular}{|c|c|c|c|c|} \hline \hline
Polymer & $N$  & $T$ [K] & $\rho$ [sites/\AA$^3$]  & $R_g$ [\AA] \\
\hline
PE      & 44   & 400     & 0.0324                  & 10.50  \\
PE      & 66   & 448     & 0.0329                  & 13.32  \\
PE      & 96   & 453     & 0.0328                  & 16.78  \\
sPP     & 96   & 453     & 0.0328                  & 13.93  \\
hhPP    & 96   & 453     & 0.0336                  & 13.54  \\
iPP     & 96   & 453     & 0.0328                  & 11.34  \\
\hline \hline
\end{tabular}
\pagebreak

TABLE II. Polyolefin blends ($T=453$ K and $N_A=N_B=96$).\\
\begin{tabular}{|c|c|c|c|c|} \hline \hline
Blend [A/B]& $\phi$ & $\rho$ [sites/\AA$^3$] & $R_{gA}$ [\AA] & $\gamma$ \\
\hline
hhPP/PE    & 0.50   & 0.0332                 & 12.32          & 1.34 \\
iPP/PE     & 0.25   & 0.0328                 & 11.35          & 1.47 \\
\hline \hline
\end{tabular}

\pagebreak
FIG.\ 1. Plot of $h(r)$ against $r/R_g$ for PE melts. Theory (full lines) is
compared with UA-MD (filled symbols) and mesoscale (open symbols) simulations
for $N = 96 \mbox{ (squares), } 66 \mbox{ (circles), and }
44 \mbox{ (diamonds)}$. Inset shows a plot of $\beta v(r)$ against $r/R_g$
for the $N = 44$ case.

FIG.\ 2. Plot of $h(r)$ against $r/R_g$ for melts of different polyolefins.
Theory (full lines) is compared with UA-MD simulations for PE (squares), sPP
(circles), iPP (downward triangles). Inset shows hhPP (upward triangles).

FIG.\ 3. Plot of $h(k)$ against $k \geq R_{g}^{-1}$. Theory (full lines) is
compared with UA-MD simulations (filled symbols) and mesoscale simulations (open
symbols). Left panel shows PE melts with
$N = 96 \mbox{ (squares), } 66 \mbox{ (circles), and } 44 \mbox{ (diamonds)}$.
Right panel displays melts of different polyolefins: PE (circles),
sPP (squares), iPP (downward triangles); inset shows hhPP (upward triangles).

FIG.\ 4. Plot of $h(r)$ against $r$ for blends. Theoretical curves in athermal
(solid lines) and thermal (dashed lines) limits are compared with $AA$-
(circles), $AB$- (diamonds) and $BB$-terms (squares) of $h(r)$ from UA-MD
simulations of representative $\phi$-symmetric (upper panel) and $\phi$-
asymmetric (lower panel) blends.

\end{document}